\documentclass{PoS}

\usepackage{subfigure}
\makeatletter
\newcommand{\figcaption}[1]{\def\@captype{figure}\caption{#1}}
\newcommand{\tblcaption}[1]{\def\@captype{table}\caption{#1}}

\title{The order of the deconfinement phase transition in a heavy quark mass region}

\ShortTitle{The order of the deconfinment phase transition}

\author{\speaker{H.~Saito}, S.~Aoki, K.~Kanaya, H.~Ohno\\
        Graduate School of Pure and Applied Sciences, University of Tsukuba, Tsukuba, Ibaraki 305-8571, Japan\\
        E-mail: \email{saitouh@het.ph.tsukuba.ac.jp}
        }
\author{S.~Ejiri\\
        Graduate School of Science and Technology, Niigata University, Niigata 950-2181, Japan
}
\author{T.~Hatsuda\\
Department of Physics, The University of Tokyo, Tokyo 113-0033, Japan
}
\author{Y.~Maezawa\\
Mathematical Physics Laboratory, RIKEN Nishina Center, Saitama 351-0198,
Japan
}
\author{T.~Umeda\\
Graduate School of Education, Hiroshima University, Hiroshima 739-8524, Japan
}
\author{(WHOT-QCD collaboration)}

\abstract{
We study the quark mass dependence of the QCD phase transition by an effective potential defined through the distribution function of observables. 
As a test of the method, we study the first order deconfinement phase transition in the heavy quark mass limit and its fate at lighter quark masses. 
We confirm that the distribution function for the plaquette has two peaks indicating that the phase transition is of first order in the heavy quark limit. 
We then study the quark mass dependence of the distribution function by a reweighting method combined with the hopping parameter expansion. 
We find that the first order transition turns into a crossover as the quark mass decreases.
We determine the critical point
 for the cases of $N_{\rm f}=1$, $2$, $3$ and $2+1$.
We find that the probability distribution function provides us with a powerful tool to study the order of transitions. 
}

\FullConference{The XXVIII International Symposium on Lattice Field Theory, Lattice2010\\
		June 14-19, 2010\\
		Villasimius, Italy}

\begin{document}

\section{Introduction}

It is of great importance to know the nature of the QCD phase transition in understanding the evolution of the early universe. 
The nature of the phase transition depends on the quark masses. 
The deconfinement phase transition is expected to be first order when the masses of up, down and strange quarks are either sufficiently large or small, 
and is crossover in the intermediate region between them.
At the physical quark masses, previous studies with staggered quarks strongly suggest that the transition is crossover \cite{Wupper2006,BNL-Bie2009}. 
A confirmation of this result by other fermion formulations such
as Wilson-type fermions or by other methods for the analysis, however,
is mandatory to draw a definite conclusion on the order of the
transition at physical quark masses.

In this report, we propose a method based on measurements of the probability distribution function of a physical quantity to identify the order of the phase transition.
Since two phases coexist at a first order phase transition point, one can identify the first order phase transition by measuring the distribution function,
which must have two peaks at the first order phase transition point 
where two phases coexist with equal probability. 
In our study we take the plaquette, i.e. $1 \times 1$ Wilson loop, to label
the state.

As a test of the distribution function method, we 
investigate the quark mass dependence of the order of the QCD phase
transition in a  large quark mass region, where numerical calculations
are much easier than those in the light quark mass region.
In addition, the reweighting \cite{Ferrenberg:1988yz,Ejiri:2007ga}  technique is used to vary quark masses.

This paper is organized as follows: Basic properties of the plaquette
distribution function are discussed in Sec.~\ref{sec:plaquette}. 
The method to calculate the plaquette distribution function by the hopping parameter expansion
in the heavy quark mass region is introduced in Sec.~\ref{sec:quarkdet}. 
In Sec.~\ref{sec:simulations}, we present results for an effective potential defined from the distribution
function, and show that the order of phase transition changes from the
first order to crossover as the quark mass decreases from infinity. We
then evaluate the location of the critical point. 
The paper is summarized in Sec.~\ref{sec:conclusion}.

\section{Probability distribution function}
\label{sec:plaquette}

The probability distribution function provides us with one of the most fundamental approaches to identify the order of the phase transition.
Because there exist two phases simultaneously at a first order phase transition point,
we expect that the probability distribution has two peaks there.
In this report, we study the distribution function of the average plaquette $P$, i.e. $1 \times 1$ Wilson loop.
We use the plaquette action $(S_g)$ for the gauge part and the standard Wilson quark action $(S_q)$ for the quark part:
\begin{eqnarray}
S_g &=& -6  \beta N_{\rm site}P, \\
S_q &=& 
\displaystyle \sum_{f=1}^{N_{\rm f}} \left\{ \sum_n\bar{\psi}_n^{(f)}\psi_n^{(f)} 
-\kappa_f \displaystyle \sum_{n,\mu}\bar{\psi}_n^{(f)} 
\left[ (1-\gamma_{\mu})U_{n,\mu}\psi_{n+\hat{\mu}}^{(f)}
+(1+\gamma_{\mu})U_{n-\hat{\mu},\mu}^{\dagger}\psi_{n-\hat{\mu}}^{(f)} \right] \right\} \\
& \equiv & \sum_{f=1}^{N_{\rm f}} \left\{ \sum_{n,m} \bar{\psi}_n^{(f)} M_{nm} (\kappa_f) 
\psi_m^{(f)} \right\},
\end{eqnarray} 
where $N_{\rm site}=N_s^3\times N_t$ is the number of sites, $N_{\rm f}$ is the number of flavors, 
$\kappa_f$ is the hopping parameter, and $\beta = 6/g^2$.
The quark mass is controlled by $\kappa_f$, which is proportional to $1/\kappa_f$ when $\kappa_f$ is small,
while the lattice spacing is mainly controlled by $\beta$.
For the case of degenerate quark masses, 
i.e. $\kappa_f=\kappa$ for $f=1, \cdots, N_{\rm f}$, 
the probability distribution function for the plaquette variable is defined by 
\begin{eqnarray}
w(P', \beta, \kappa) 
& = & \int {\cal D} U \ \delta(P'-P[U]) \ (\det M(\kappa))^{N_{\rm f}} 
e^{6\beta N_{\rm site} P},
\label{eq:pdist}
\end{eqnarray}
where $\delta(x)$ is the delta function. 
The partition function is given by 
${\cal Z} (\kappa, \beta) = \int w(P', \beta, \kappa) dP' $.
In the followings, we denote $P'$ simply as $P$.

The plaquette distribution function can be obtained by the histogram of $P$ and has the following useful property:
Under the parameter change from $\beta_0$ to $\beta$, the weight $w(P, \beta,\kappa)$ becomes
\begin{equation}
w(P,\beta,\kappa)=e^{6(\beta-\beta_0)N_{\rm site}P}w(P,\beta_0,\kappa).
\end{equation}
The effective potential defined by
\begin{equation}
V(P,\beta,\kappa) = -\ln w(P,\beta,\kappa)
\end{equation}
varies as
\begin{equation}
V(P, \beta, \kappa) 
= V(P, \beta_0, \kappa) - 6(\beta - \beta_0)N_{\rm site}P
\end{equation}
under this parameter change.
From this property, we find that 
$d V/dP$ changes trivially as
\begin{eqnarray}
\frac{d V}{dP} (P,\beta, \kappa)
= \frac{d V}{dP} (P,\beta_0, \kappa) -6 (\beta - \beta_0) N_{\rm site},
\label{eq:derrewbeta}
\end{eqnarray}
and $d^2 V/dP^2$ is unchanged under the $\beta$ shift.
Therefore, the shape of $dV/dP$ as a function of $P$ does not change 
with $\beta$ up to the $\beta$-dependent constant.

When a first order transition exists, the distribution is a double-peaked function at the transition point. 
The effective potential is then a double-well function and the derivative of $V$ is an S-shaped function. 
In the transition region, $d V/dP$ vanishes at three points. 
To locate the transition point, the fine turning of $\beta$ to the transition point is not required because the $P$-dependence of $d V/dP$ itself is independent of $\beta$. 
Therefore, the measurement of $d V/dP$ is useful to identify the first order phase transition. 

Next, we discuss the $\kappa$-dependence of $V$ considering the ratio of the distribution functions at different $\kappa$ and $\kappa_0$:
\begin{eqnarray}
R(P', \kappa, \kappa_0) 
 \equiv& \frac{w(P', \beta, \kappa)}{ w(P', \beta, \kappa_0)} 
= \frac{\int {\cal D} U \delta(P'-P[U]) (\det M(U, \kappa))^{N_{\rm f}}}{
\int {\cal D} U \delta(P'-P[U]) (\det M(U, \kappa_0))^{N_{\rm f}}} 
= \frac{ \left\langle \delta(P'-P[U]) 
\frac{(\det M(U, \kappa))^{N_{\rm f}}}{(\det M(U, \kappa_0))^{N_{\rm f}}} 
\right\rangle_{(\beta, \kappa_0)} }{
\left\langle \delta(P'-P[U]) \right\rangle_{(\beta, \kappa_0)}}. 
\label{eq:rkdef}
\end{eqnarray}
This $R(P, \kappa, \kappa_0)$ is independent of $\beta$.
Using $R(P, \kappa, \kappa_0)$, the $\kappa$-dependence of the effective potential is given by the following equation:
\begin{equation}
V(P, \beta, \kappa) 
= -\ln R(P, \kappa, \kappa_0) + V (P,\beta, \kappa_0), 
\label{eq:veff}
\end{equation}
under the change from $\kappa_0$ to $\kappa$.

This argument can be generalized to improved gauge actions easily. 
For the case of improved gauge actions including larger Wilson loops, 
we should define the average plaquette as $P=-S_g/(6N_{\rm site}\beta)$, 
which is a linear combination of Wilson loops. 
On the other hand, $\beta$-dependent improved quark actions make the analysis more complicated.

\section{Quark determinant in the heavy quark mass region}
\label{sec:quarkdet}
To investigate the quark mass dependence of the plaquette effective potential, 
we evaluate the quark determinant by the Taylor expansion with respect to the hopping parameter $\kappa$ in the vicinity of the simulation point $\kappa_0$:
\begin{eqnarray}
\ln \left[ \frac{\det M(\kappa)}{\det M(\kappa_0)} \right]
= \sum_{n=1}^{\infty} \frac{1}{n!} 
\left[ \frac{\partial^{n} (\ln \det M)}{\partial \kappa^{n}} 
\right]_{\kappa_0} (\kappa - \kappa_0)^{n} 
= \sum_{n=1}^{\infty} \frac{1}{n!} {\cal D}_{n} (\kappa - \kappa_0)^{n} , 
\label{eq:tayexp}
\end{eqnarray}
where 
\begin{eqnarray}
{\cal D}_n \equiv \left[ \frac{\partial^n (\ln \det M)}{\partial \kappa^n} \right]_{\kappa_0}
= (-1)^{n+1} (n-1)! \ {\rm tr} 
\left[ \left( M^{-1} \frac{\partial M}{\partial \kappa} \right)^n \right]_{\kappa_0}. 
\label{eq:derkappa}
\end{eqnarray}
Calculating the derivative of the quark determinant ${\cal D}_{n}$,
the $\kappa$-dependence of the effective potential can be estimated.

In this study, we focus on the boundary which separates the first order transition region near the quenched limit and the crossover region.
The boundary is expected to exist near $\kappa=0$.
We adopt $\kappa_0=0$ where $M_{x,y} = \delta_{x,y}$.
The $(\partial M/\partial \kappa)_{x,y}$ is the gauge connection between $x$ and $y$, 
and the nonzero contribution of ${\cal D}_{n}$ is given by Wilson loops and Polyakov loops. 
Considering the anti-periodic boundary condition and gamma matrices in the hopping terms, 
the leading contributions of the Taylor expansion are given by the $\kappa^4$ term and $\kappa^{N_t}$ term:
\begin{eqnarray}
\ln \left[ \frac{\det M (\kappa)}{\det M (0)} \right]  
= 288 N_{\rm site} \kappa^4 P + 12\times 2^{N_t}N_s^3 \kappa^{N_t} {\rm Re} \Omega 
+ \cdots , 
\label{eq:detm}
\end{eqnarray}
where $\Omega$ is the Polyakov loop:  
\begin{equation}
\Omega = \frac{1}{N_s^3}
\displaystyle \sum_{\mathbf{n}} \frac{1}{3} {\rm tr}\left[ 
U_{\mathbf{n},4} U_{\mathbf{n}+\hat{4},4} U_{\mathbf{n}+2\hat{4},4} \dots U_{\mathbf{n}+(N_t-1)\hat{4},4} \right].
\label{eq:ploop}
\end{equation}
The ratio $R(P, \kappa, 0) = w(P, \beta, \kappa) / w(P, \beta, 0)$ 
is calculated by the following equation in the region of small $\kappa$ for any $\beta$:
\begin{eqnarray}
R(P', \kappa, 0) 
&=& e^{N_{\rm f} N_{\rm site} 288 \kappa^4 P' }
\frac{ \left\langle \delta(P'-P[U]) \exp [N_{\rm f} N_s^3
(12\times 2^{N_t} \kappa^{N_t} {\rm Re}\Omega[U] 
+ \cdots ) ] \right\rangle_{(\beta, \kappa_0=0)} }{
\left\langle \delta(P'-P[U]) \right\rangle_{(\beta, \kappa_0=0)}}. 
\label{eq:rewquench}
\end{eqnarray}
For $N_t=4$ the truncation error is $O(\kappa^6)$.
The contribution from the plaquette can be absorbed in the redefinition of $\beta$.
Using Eqs.~(\ref{eq:veff}) and (\ref{eq:rewquench}), we investigate 
the $\kappa$-dependence of the effective potential in the heavy quark mass region.

\section{Numerical simulations and the results}
\label{sec:simulations}
In the heavy quark mass limit we perform simulations of SU(3) pure gauge theory on a $24^3 \times 4$ lattice.
To generate the configurations, the pseudo heat bath algorithm of SU(3) gauge theory is used, 
and the over relaxation is performed 4 times every update.
Because the effective potential must be investigated in a wide range of the plaquette value to identify the order of the phase transition, 
we performed simulations at five points in the range of $\beta = 5.68$ -- $5.70$. 
In the calculation, we approximate the delta function by a Gaussian function:
$\delta(x) \approx 1/(\Delta \sqrt{\pi})\exp{ \left[-(x/\Delta)^2\right] }$.
Examining the resolution and the statistical error in the distribution function, 
we adopt $\Delta = 0.000283$.

\begin{figure}[t] 
   \centering
   \begin{minipage}{7cm}
   \includegraphics[width=6.8cm, clip]{./combine_v3.eps} 
   \caption{Derivatives of the effective potential for pure gauge at $\beta=5.68$--$5.70$. 
   The black line is the average.
}
   \label{fig:combine}
  \end{minipage}\ \ \ \ \ \ 
  \begin{minipage}{7cm}
    \includegraphics[width=9cm, height=6cm, keepaspectratio, trim=30 25 0 45, clip]{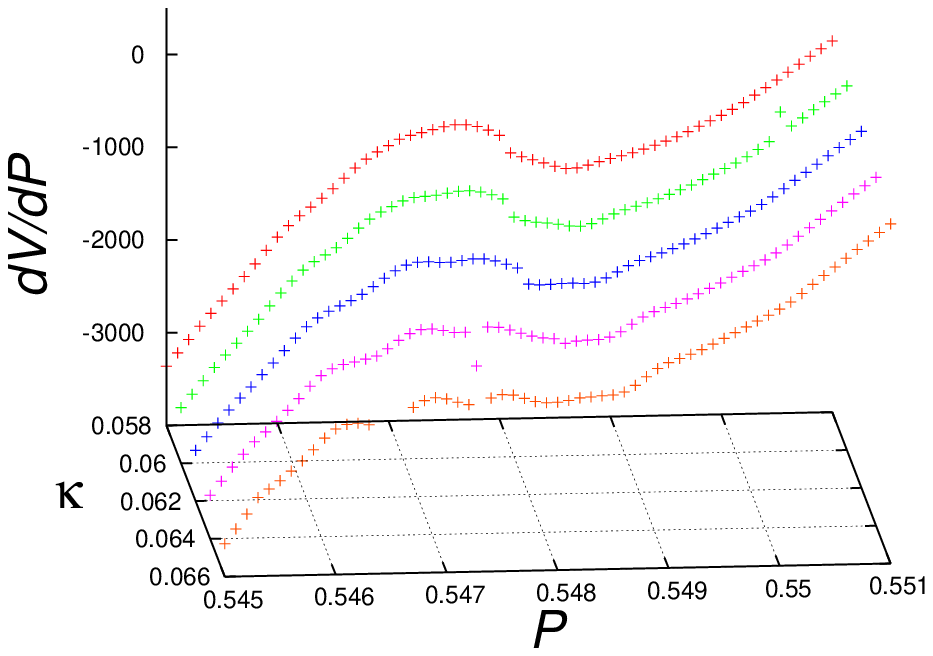}
     \caption{The $\kappa$-dependence of the derivative of the effective potential at $\beta=5.69$.}
     \label{fig:dV}
   \end{minipage}
\end{figure}

We then calculate the derivative $dV/dP$ by the difference between the potentials at $P$ and $P + \Delta P$. 
We adopt $\Delta P=0.0001$ with which the $\Delta P$-dependence is much smaller than the statistical error.
The results of $dV/dP$ at $\kappa=0$ are shown in Fig.~\ref{fig:combine}. 
In this figure, we adjust results at different $\beta$'s to $\beta=5.69$ by using Eq.~(\ref{eq:derrewbeta}).
The results of $dV/dP$ obtained in simulations at different $\beta$ is consistent within errors, 
though the ranges of $P$ in which $V$ is reliably obtained are different.
The jackknife method is used to estimate the statistical error of the effective potential and its derivatives.
We combine these data obtained by an weighted average with the inverse-square of each error,
which is shown by a black line in Fig.~\ref{fig:combine}.
The data is excluded from the average over different ensembles if the error is large and $P$ is far away from the peak of the distribution
function at each $\beta$.
We find, from this figure, that $dV/dP$ at $\kappa=0$ is not a monotonically increasing function, 
meaning the double-well structure of the effective potential.

The quark mass dependence of the effective potential is investigated by calculating $R(P, \kappa, 0)$ at the order of $\kappa^4$  in Eq.~(\ref{eq:rewquench}). 
Using the data of $w(P,\beta, \kappa=0)$ and $R(P, \kappa, 0)$, we evaluate the $dV/dP$ up to a finite value of $\kappa$. 
The results for $N_{\rm f}=2$ are plotted in Fig.~\ref{fig:dV}. 
The S-shape structure becomes milder as $\kappa$ increases and seems to turn into a monotonically increasing function around $\kappa = 0.066$. 
This behavior suggests that the first order phase transition at $\kappa=0$ becomes weaker as $\kappa$ increases and the transition changes to crossover at $\kappa \approx 0.066$ or larger. 

To evaluate more precisely the value of $\kappa$ at the boundary where the first order phase transition is terminated $(\kappa_{\rm cp})$,
we calculate the second derivative of $V(P, \beta, \kappa)$ by a numerical differentiation of $dV/dP$.
When the transition is first order, there is a region where the second derivative of $V(P, \beta, \kappa)$ is negative between the two bottoms of $V(P, \beta, \kappa)$. 
The first order transition region is thus identified by measuring the sign of the second derivative of $V$. 
As discussed in Sec.~\ref{sec:plaquette}, the second derivative of $V(P, \beta, \kappa)$ is independent of $\beta$. 
Therefore, the identification can be performed without considering the $\beta$-dependence.

\begin{figure}[t]
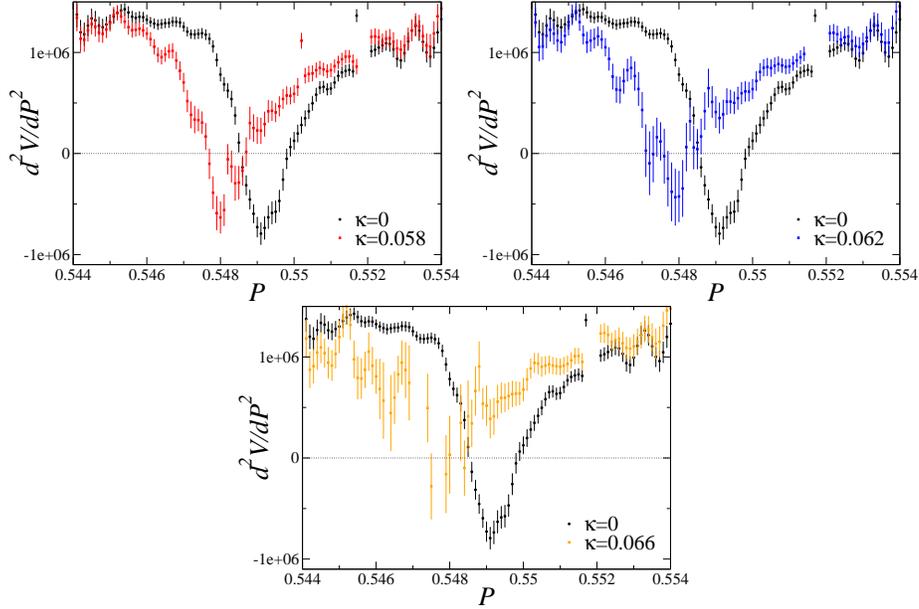
 
  \begin{minipage}{6cm}
      \centering
    \includegraphics[width=6cm,  clip]{./ddV_K0058.eps}
   \end{minipage}
   \begin{minipage}{6cm}
       \centering
   \includegraphics[width=6cm, clip]{./ddV_K0062.eps}
   \end{minipage}
   \centering
   \begin{minipage}{6cm}
     \includegraphics[width=6cm, clip]{./ddV_K0066.eps}
   \end{minipage}
    \caption{The second derivative of the effective potential at $\kappa=0.058, 0.062, 0.066$. The black symbols are the results at $\kappa=0$.}
    \label{fig:3d_ddV}
\end{figure}

\begin{figure}[t]
    \begin{minipage}{6.5cm}
    \centering
      \includegraphics[width=6.5cm, trim=0 0 0 0, clip]{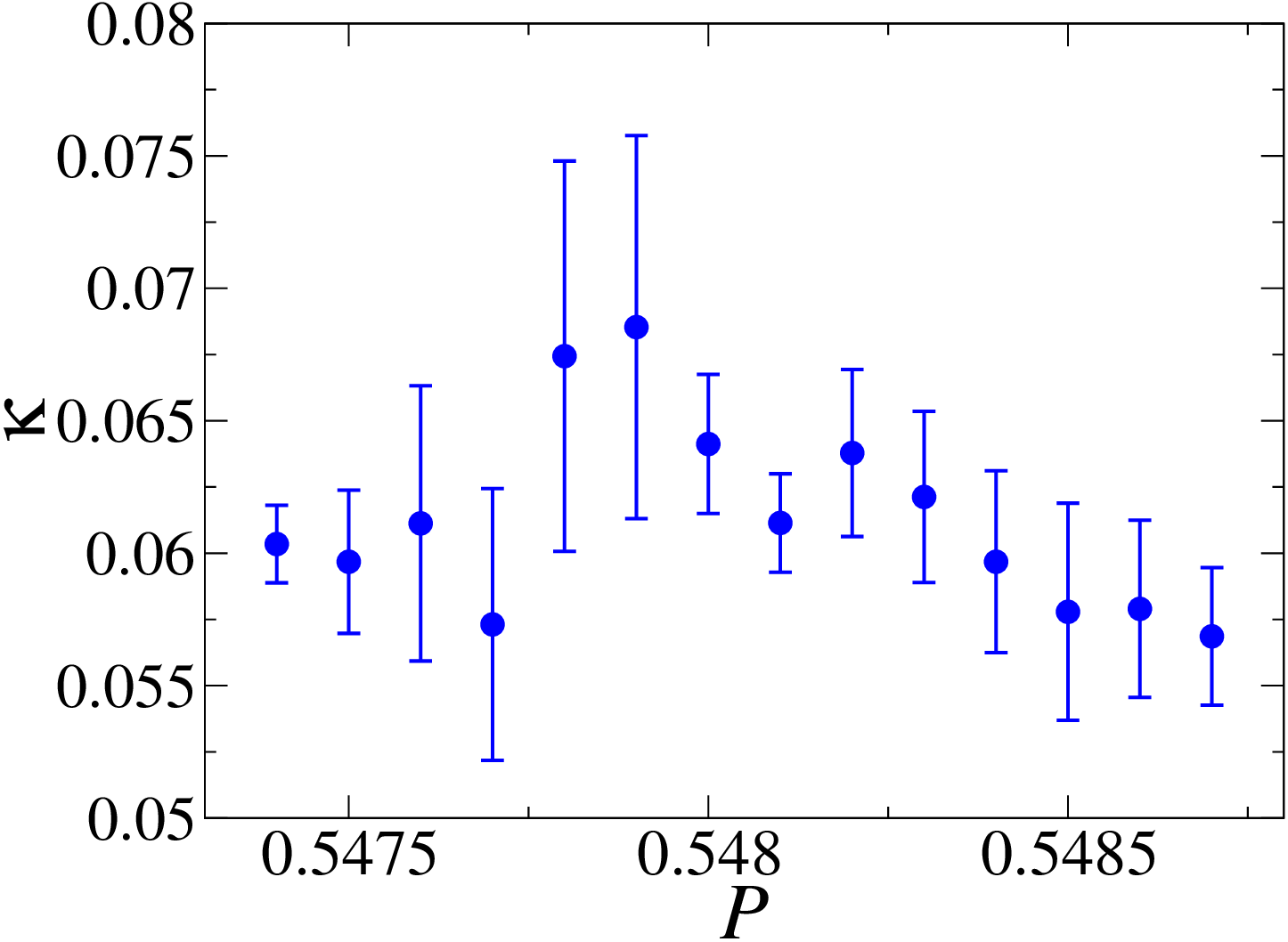} \ 
   \caption{The $\kappa$ value at which $d^2V/dP^2=0$ for each fixing $P$ for $N_{\rm f}=2$.}
   \label{fig:Kep_search}
   \end{minipage}\ \ \ \ \ 
  \centering 
   \begin{minipage}{8cm}
    \centering
     \includegraphics[width=6cm, trim=3 0 3 0, clip]{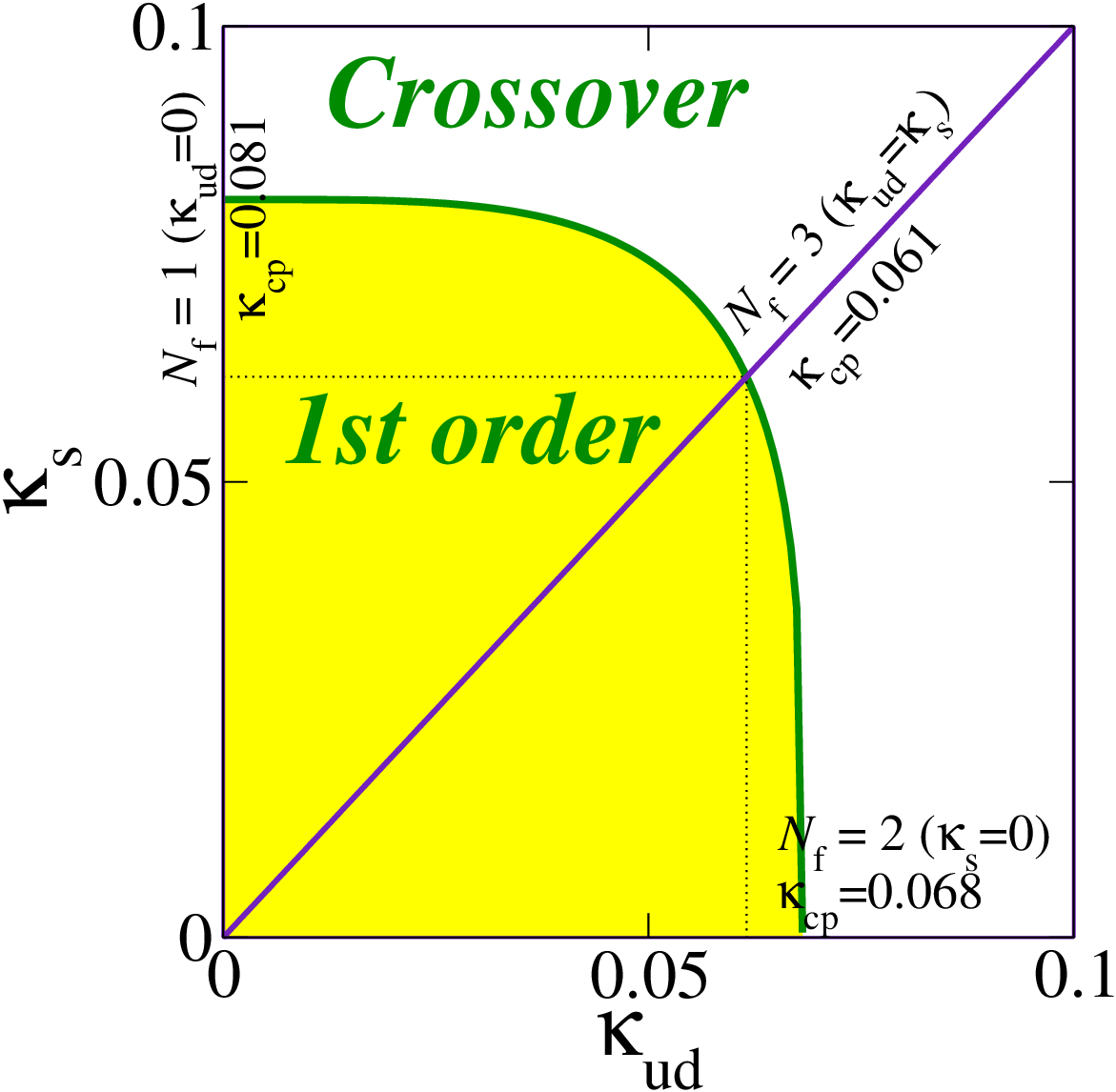}
     \caption{The quark mass dependence of the order of phase transition in $2+1$ flavor QCD.}
     \label{fig:Nf_Kep}
   \end{minipage}
\end{figure}

We plot the results of $d^2 V/dP^2$ at $\kappa=0.058$, $0.062$ and $0.066$ for $N_{\rm f}=2$ in Fig.~\ref{fig:3d_ddV}. 
The range of $P$ where $d^2 V/dP^2 < 0$ becomes narrower  at non-zero $\kappa$ than that at $\kappa=0$. 
We calculate $d^2 V/dP^2$ varying $\kappa$ with fixing $P$, and find the value of $\kappa$ at which $d^2 V/dP^2 =0$. 
The result is plotted in Fig.~\ref{fig:Kep_search} for $N_{\rm f}=2$. 
In the region below the symbols, the curvature of $V$ is negative.
Fig.~\ref{fig:Kep_search} shows a peak around $P=0.5478$, and $d^2 V/dP^2$ is positive for all $P$ when $\kappa$ is larger than the peak height $0.068(7)$. 
This means that the phase transition is no longer first order above the critical value of $\kappa_{\rm cp}=0.068(7)$ for $N_{\rm f}=2$. 
We calculate the maximum value of $\kappa$ where $d^2 V/dP^2 =0$ , which is $\kappa_{\rm cp}$, with estimating the jackknife error.
The error is estimated by a jackknife method.

This analysis can be also applied to $N_{\rm f}=2+1$ QCD having different light quark mass and strange quark mass as well as $N_{\rm f}=1$ and $3$ cases. 
For the case of degenerate quark masses, the second derivative of $V$ does not change if the term $N_{\rm f}\kappa^{N_t}$ is constant.
From this property and the result of the critical point for $N_{\rm f}=2$, we find $\kappa_{\rm cp}=0.081(8)$ for $N_{\rm f}=1$ and  $0.061(6)$ for $N_{\rm f}=3$.
The result of $N_{\rm f}=1$ is consistent with the results obtained by an effective $Z(3)$ model in Ref.~\cite{Alexandrou:1998wv}.
For the case of $N_{\rm f}=2+1$, 
within the leading order of the hopping parameter expansion, the quark determinant in the partition function is given by 
\begin{eqnarray}
\ln \left[ \frac{(\det M (\kappa_{\rm ud}))^2 \det M (\kappa_{\rm s})}{(\det M (0))^3} \right]  
=  288 N_{\rm site}(2 \kappa_{\rm ud}^4 +\kappa_{\rm s}^4) P + 12 \times 2^{N_t} N_s^3(2 \kappa_{\rm ud}^{N_t} + \kappa_{\rm s}^{N_t}) {\rm Re}\Omega 
+ \cdots , 
\label{eq:detm2+1}
\end{eqnarray}
where $\kappa_{\rm ud}$ and $\kappa_{\rm s}$ are hopping parameters for light and strange quarks. 
Because the contribution from the plaquette term in this equation does not affect the second derivative of $V$, 
the difference from the case of $N_{\rm f}$ is just the replacement from $N_{\rm f} \kappa^{N_t}$ to $2 \kappa_{\rm ud}^{N_t} + \kappa_{\rm s}^{N_t}$. 
Thus the line which separates the first order phase transition and the crossover is given by 
\begin{equation}
2 \kappa_{\rm ud}^{N_t} + \kappa_{\rm s}^{N_t} =\left( \kappa_{\rm cp}^{N_{\rm f}=1}\right)^{N_t}, 
\end{equation}
where $N_t=4$ and $\kappa_{\rm cp}^{N_{\rm f}=1}=0.081(8)$ in this study.
We draw the line in Fig.~\ref{fig:Nf_Kep}.

\section{Conclusion}
\label{sec:conclusion}
We studied the order of the deconfinement phase transition in the heavy quark mass region of QCD by calculating the probability distribution function of the average plaquette.
The distribution function at $\kappa=0$ is evaluated by measuring the histogram in quenched QCD, and the $\kappa$-dependence is investigated by using the reweighting method within the approximation by the leading approximation of the hopping parameter expansion for $\ln \det M$. 
Because the distribution function must be measured in a wide range of $P$ to study the order of the phase transition, we performed quenched simulations at 5 points of $\beta$ and combined these results.

We found that the distribution function shows two peaks at $\kappa=0$ indicating the first order transition, and the double-peaked shape becomes weaker as $\kappa$ increases. 
We calculated the critical point $\kappa_{\rm cp}$. 
The results are $\kappa_{\rm cp}=0.081(8)$, $0.068(7)$ and $0.061(6)$ for $N_{\rm f}=1$, $2$ and $3$, respectively. 
The boundary of the first order transition region for $N_{\rm f}=2+1$ QCD is also determined in the two-dimensional parameter space of $\kappa_{\rm ud}$ and $\kappa_{\rm s}$.

\vspace{5mm}
This work is in part supported 
by Grants-in-Aid of the Japanese Ministry of Education, Culture, Sports, Science and Technology, 
(Nos.  21340049, 22740168, 22840020, 20340047)
and by the Grant-in-Aid for Scientific Research on Innovative Areas
(Nos. 20105001, 20105003).

\end{document}